\documentclass[a4paper,11pt]{article}
\usepackage{pos}
\usepackage[inline]{enumitem}
\usepackage{color}
\definecolor{cRed}{RGB}{196,0,100}

\title{Indirect bounds on new physics for $R(D^{(*)})$}

\author*[a]{Benjam\'\i{}n Grinstein}
\author[b]{Jason Aebischer}

\affiliation[a]{Department of Physics, UC San Diego\\
 9500 Gilman Dr, La Jolla, CA, USA}

\affiliation[b]{Physik-Institut, Universit\"at Z\"urich\\
  CH-8057 Z\"urich, Switzerland}

\emailAdd{bgrinstein@ucsd.edu}
\emailAdd{jason.aebischer@physik.uzh.ch}

\abstract{The Standard Model prediction of the $B_c$ lifetime is discussed, together with the dominant uncertainties and strategies on how to improve them. Furthermore, a new method to compute the $B_c$ lifetime based on the operator product expansion is proposed. It relies on differences of $B,\,D$ and $B_c$ meson decay rates, in which the free-quark contributions cancel out, reducing the uncertainty of the theory prediction.}

\FullConference{%
  11th International Workshop on the CKM Unitarity Triangle (CKM2021)\\
  22-26 November 2021\\
  The University of Melbourne, Australia
}

\begin{document}
\maketitle

\section{Introduction}

A precisely calculated $B_c$ meson lifetime puts stringent constraints
on New Physics models containing new scalars, for example, scalar Leptoquarks and Two-Higgs-Doublet models
\cite{Alonso:2016oyd,Blanke:2018yud}. These are interesting
insofar they explain anomalies reported in $R(D)$ and $R(D^*)$ measurements.

Beyond these anomalies, the $B_c=(\overline b c)$ meson is an
interesting particle to study, since it contains two different heavy
quarks. We expect it can be well described with
Non-Relativistic QCD (NRQCD), in which an expansion in the heavy quark
velocities $v_b$ and $v_c$ is carried out. Together with the
operator production expansion (OPE) approach this has lead to the most
precise theory prediction of the $B_c$ lifetime
\cite{Beneke:1996xe,Bigi:1995fs,Chang:2000ac}. Other, less systematic approaches
include QCD Sum Rules~\cite{Kiselev:2000pp} as well as Potential
models \cite{Gershtein:1994jw}, which lead to comparable results.

From the experimental point of view the lifetime of the $B_c$ is very precisely measured to be
%
\(
\tau_{B_c}^{\text{exp}} = 0.510(9)\text{ps}
\)
(averaged value of the LHCb \cite{LHCb:2014ilr,LHCb:2014glo} and CMS
\cite{CMS:2017ygm} measurements), that is,
\begin{equation}\label{eq:Gexp}
  \Gamma_{B_c}^\text{exp} = 1.961(35) \,\text{ps}^{-1}\,,
\end{equation}
%
%
for the corresponding total decay rate.
This precision is however not matched by {the theory prediction due to}
large uncertainties {in the calculation}
 that arise largely from neglecting higher order
non-perturbative corrections, parametric uncertainties and
difficulties accounting for the strange quark mass, among others. The
main uncertainties stem however from the treatment of the masses of
the quarks inside the $B_c$, and are inextricably tied to the
perturbative expansion. To examine this in more detail, we studied in
Refs.~\cite{Aebischer:2021ilm,Grinstein:2022awh} three different mass
schemes in the $B_c$ decay rate in the OPE approach; these, as well as
the other sources of uncertainty, are discussed below.

\section{Mass schemes}\label{sec:mass}

\subsection{$\overline{\text{MS}}$ scheme}

In the $\overline{\text{MS}}$ mass-scheme the on-shell (OS) masses of the $\overline{b}$ and $c$ quarks are expressed in terms of the renormalized $\overline{\text{MS}}$ masses via the following equation:
\begin{equation}
  \label{eq:poleM1loop}
  m_q=\overline{m}_q(\mu)\left[1+\frac{\alpha_s(\mu)}{\pi}\left(\frac{4}{3}-\ln\left(\frac{\overline{m}_q(\mu)^2}{\mu^2}\right)\right)\right]+\mathcal{O}(\alpha_s^2)\,.
\end{equation}
%
In our computation we use the lattice results \cite{Bazavov:2018omf,Colquhoun:2014ica,Lytle:2018evc} for the $\overline{\text{MS}}$ masses, which lead to the following decay rate of the $B_c$:
\begin{equation}\label{eq:MSbar}
  \Gamma^{\overline{\text{MS}}}_{B_c} = (1.51\pm 0.38|^{\mu}\pm 0.08|^{\text{n.p.}}\pm 0.02|^{\overline{m}}  \pm0.01|^{m_s}\pm 0.01|^{V_{cb}})\,\,\text{ps}^{-1}\,,
\end{equation}
where the third uncertainty is due to the $\overline{\text{MS}}$
masses. The other uncertainties will be discussed in the following
section. The value in \eqref{eq:MSbar} is to be compared with
the experimental value, Eq.~\eqref{eq:Gexp}.

\subsection{Upsilon scheme}

In this mass scheme, the OS mass of the $\overline b$ quark is expressed in terms of the very precisely measured Upsilon 1S state, by using the relation \cite{Pineda:1997hz,Melnikov:1998ug}

\begin{equation}
  \frac{\tfrac12m_{\Upsilon}}{m_b}=1-\frac{(\alpha_s C_F)^2}{8}
  \left\{1
    +\frac{\alpha_s}{\pi}\left[\left(\ln\left(\frac{\mu}{\alpha_sC_Fm_b}\right)+\frac{11}{6}\right)\beta_0-4\right]^2+\cdots\right\}\,,
\end{equation}
where $\beta_0$ is the one-loop beta function factor of the strong
coupling constant. A similar relation is used to express the charm
quark mass in terms of the $J/\Psi$ mass. We use the
PDG values $m_{\Upsilon}=9460.30(26)$ MeV and $m_{J/\Psi}=3096.900(6)$
MeV \cite{Tanabashi:2018oca}, which gives a $B_c$ decay rate of
\begin{equation}\label{eq:upsG}
  \Gamma^{\text{Upsilon}}_{B_c} = (2.40\pm 0.19|^{\mu}\pm 0.21|^{\text{n.p.}} \pm0.01|^{m_s}\pm 0.01|^{V_{cb}})\,\,\text{ps}^{-1} \,,
\end{equation}
where the uncertainties of $m_{\Upsilon}$ and $m_{J/\Psi}$ are completely negligible.

\begin{figure}[t]
  \begin{center}
    \includegraphics[width=0.45\textwidth]{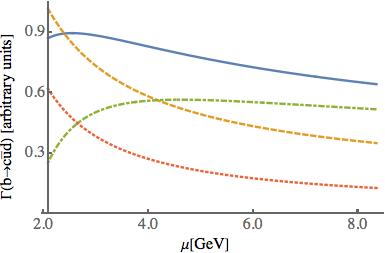}\hspace{0.05\textwidth}
    \includegraphics[width=0.45\textwidth]{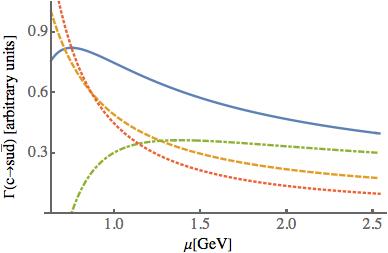}
    \caption{\label{fig:mu-dep} Scale dependence of the LO decay rates $\Gamma(b\to cud)$ (left panel) and
      $\Gamma(c\to sud)$ (right panel) in the $\overline{\text{MS}}$ scheme. The NLO (solid-blue) and LO (dashed-orange) calculations are shown, respectively. The LO calculation to which the term with
      the explicit factor of $\alpha_s\ln(\mu)$ in the NLO decay rate is added is shown in green, displaying cancellation of
      scale dependence at $\mathcal{O}(\alpha_s)$. The NLO decay rate omitting the term with an explicit factor of
      $\ln(\mu)$ is given by the dotted-red line.}
    \end{center}
\end{figure}

\subsection{Meson scheme}

As a third scheme we use the so-called meson scheme, where the OS quark masses are expressed in terms of the meson masses by use of the HQET relation
\begin{equation}
  \label{eq:poleMassDiff}
  m_b-m_c=\overline{m}_B-\overline{m}_D+\frac12\lambda_1\left(\frac1{m_b}-\frac1{m_c}\right)
  +\cdots
\end{equation}
%
where $\lambda_1=-0.27 \pm 0.14$ \cite{Hoang:1998ng}, and $\overline{m}_B=\frac14(3 m_{B^*}+m_B)$ and $\overline{m}_D=\frac14(3 m_{D^*}+m_D)$ denote the spin and isospin-averaged meson masses. In this scheme we obtain

\begin{equation}\label{eq:mesonG}
  \Gamma^{\text{meson}}_{B_c} = (1.70\pm 0.24|^{\mu}\pm 0.20|^{\text{n.p.}} \pm0.01|^{m_s}\pm 0.01|^{V_{cb}})\,\,\text{ps}^{-1} \,,
\end{equation}
where the obtained value is in rather good agreement with the measurement in Eq.~\eqref{eq:Gexp}.

\section{Uncertainties}\label{sec:uncertainties}


\subsection{Scale dependence}

The residual renormalization-scale dependence from truncating the loop expansion is the largest uncertainty in the $B_c$ lifetime. It enters mainly through the OS mass replacements of the quarks in the three different schemes, since these relations are only used at the one-loop level. The scale dependence is largest in the $\overline{\text{MS}}$ scheme, which is illustrated in Fig.~\ref{fig:mu-dep}: It depicts the scale dependence of the leading order (LO) quark decay rates $\Gamma(b\to cud)$ and $\Gamma(c\to sud)$.

To reduce the scale dependence in our results, higher order QCD
corrections have to be incorporated in the calculation, both in the
OS mass relations and  in free-quark decay rates.

\subsection{Non-perturbative uncertainties}

Further uncertainties result from the NRQCD expansion in the quark velocities $v_b$ and $v_c$, which has been truncated at $\mathcal{O}(v^4)$. Furthermore, the non-perturbative (n.p.) parameters also have uncertainties which are incorporated in the n.p. uncertainty estimations in eqs.~\eqref{eq:MSbar}, \eqref{eq:upsG} and \eqref{eq:mesonG}. The main improvement in these uncertainties would be to include higher-order corrections in the velocity expansion. It would however also be favourable to have lattice results available for the n.p. parameters.

\subsection{Parametric uncertainties}

Additional uncertainties result from all the parameters that are involved in the calculation, the largest one stemming from the uncertainty of the CKM matrix element $V_{cb}$ given in the last uncertainties of eqs.~\eqref{eq:MSbar}, \eqref{eq:upsG} and \eqref{eq:mesonG}. In the $\overline{\text{MS}}$ scheme also the $\overline{\text{MS}}$-masses introduce a rather large uncertainty, which is shown in third uncertainty in Eq.~\eqref{eq:MSbar}.

\subsection{Strange quark mass}

 In the spectator $c$-decays a non-vanishing strange quark mass reduces the decay rate by about 7\% in the three different mass schemes. The introduced uncertainty when neglecting $m_s$ in the $\bar b$-quark decay can be estimated naively by considering the factor $(m_c/m_b)^2\sim0.1$, multiplied by the corresponding decay rate and a factor of 7\%. In the $c$-quark decays the parametric uncertainty resulting from $\overline{m}_s(2\;\text{GeV})$ leads to an uncertainty of $\Delta\Gamma_c\sim0.01\;\text{ps}^{-1}$.

\begin{table}[b]
\centering
 \begin{tabular}{|l |c |c |c |c|}
 \hline
 & $B^0,D^0$ & $B^+,D^0$ & $B^0,D^+$ & $B^+,D^+$ \\ [0.5ex]
 \hline \hline
   $\Gamma^{\text{meson}}_{B_c}$& 3.03 $\pm$ 0.54 & 3.04 $\pm$ 0.54 & 3.38 $\pm$ 0.98 & 3.39 $\pm$ 0.99 \\
  \hline
$\Gamma^{\overline{\text{MS}}}_{B_c}$ & 2.97 $\pm$ 0.42 & 2.98 $\pm$ 0.40 & 3.19 $\pm$ 0.80 & 3.19 $\pm$ 0.82
\\
 \hline
 \end{tabular}
 \caption{\small
Results obtained using the novel approach discussed in sec.~\ref{sec:newmethod} in the meson and $\overline{\text{MS}}$ scheme, using four different combinations of $B$ and $D$ mesons.
}
  \label{tab:res}
\end{table}

\section{Novel determination of $\Gamma_{B_c}$}\label{sec:newmethod}

To reduce the rather large uncertainties in the theory prediction, which mainly result from the scale dependence, we will adopt a novel method to compute the $B_c$ decay rate, first described in \cite{Aebischer:2021eio}. The idea is to make use of the non-perturbative expansion of the decay rate not only for the $B_c$ meson, but also for the $B$ and $D$ mesons, by considering the combination
\begin{align}\label{eq:diff}
  \Gamma(B)+\Gamma(D)-\Gamma(B_c) &= \Gamma^{n.p.}(B)+\Gamma^{n.p.}(D)-\Gamma^{n.p.}(B_c) \nonumber \\
  &+\,\Gamma^{\text{WA}+\text{PI}}(B)+\Gamma^{\text{WA}+\text{PI}}(D)-\Gamma^{\text{WA}+\text{PI}}(B_c)\,,
\end{align}
%
where the rates on the left-hand side are given by
\begin{equation}\label{eq:GM}
  \Gamma(H_Q) = \Gamma_Q^{(0)}+\Gamma^{n.p.}(H_Q)+\Gamma^{\text{WA}+\text{PI}}(H_Q)+\mathcal{O}(\frac{1}{m_Q^4})\,,
\end{equation}
%
for a meson $H_Q$ with heavy quark $Q$ and where WA and PI stand for Weak Annihilation and Pauli Interference contributions. On the right-hand side of Eq.~\eqref{eq:diff} the LO quark decay rates $\Gamma^{(0)}_Q$ drop out, since they are independent of meson states. Therefore the largest source of scale dependence vanishes, which reduces the error of the result. In order to determine the $B_c$ decay rate Eq.~\eqref{eq:diff} can be applied for either charged or neutral $B$ and $D$ mesons, resulting in four different ways to compute $\Gamma(B_c)$. The results using these four different combinations are given in Tab.~\ref{tab:res}.

The results from this novel approach are in tension with the experimental result in Eq.~\eqref{eq:Gexp}. Several reasons can be put forward to explain this disparity:
\begin{enumerate*}
  \item The uncertainties from NLO corrections to Wilson coefficients and free quark decay rates might be underestimated;
  \item Eye-graph contributions, neglected in lattice computations of matrix elements that we use~\cite{Becirevic:2001fy}, but estimated to be small using HQET sum rules \cite{King:2021jsq};
  \item Unexpectedly large contributions from higher dimension operators in the $1/m_Q$ expansion~\cite{King:2021xqp};
  \item Violation of quark-hadron duality.
\end{enumerate*}
%
A thorough analysis of these is in order to determine the reason for the discrepancy between the results and experiment.

\section{Summary}\label{sec:summary}

We have presented an updated analysis of the $B_c$ decay rate, following the OPE approach. Three different mass schemes have been studied, which all lead to results in agreement with experiment and with each other. Furthermore an analysis of the theory uncertainties has been presented, where the scale-dependence makes up most of the total uncertainty.

We discussed a novel method to determine $\Gamma_{B_c}$ based on
differences of $B,\,D$ and $B_c$ decay rates that allows to reduce the
scale-dependence uncertainty. The results deviate significantly from
the experimental value, and we presented various possible reasons for
this discrepancy.

\acknowledgments
Work of BG  supported in part by the U.S. Department of Energy
Grant No.~DE-SC0009919.

\small

\end{document}